\documentclass{ptephy_v1}

\usepackage{amsmath}
\usepackage{aas_macros}
\usepackage{braket}
\usepackage{graphicx}

\begin{document}

\title{Analytic solutions for neutrino-light curves of core-collapse supernovae}


\author[1,2,*]{Yudai Suwa}
\affil{Department of Astrophysics and Atmospheric Sciences, Faculty of Science, Kyoto Sangyo University, Kyoto 603-8555, Japan}
\affil[2]{Center for Gravitational Physics, Yukawa Institute for Theoretical Physics, Kyoto University, Kyoto 606-8502, Japan
\email{suwa@yukawa.kyoto-u.ac.jp}}

\author[3]{Akira Harada}
\affil[3]{Institute for Cosmic Ray Research, the University of Tokyo, Chiba 277-8582, Japan}

\author[4]{Ken'ichiro Nakazato}
\affil[4]{Faculty of Arts and Science, Kyushu University, Fukuoka 819-0395, Japan}

\author[5]{Kohsuke Sumiyoshi}
\affil[5]{National Institute of Technology, Numazu College, Shizuoka 410-8501, Japan}



\begin{abstract}%
Neutrinos are a guaranteed signal from supernova explosions in the Milky Way, and a most valuable messenger that can provide us with information about the deepest parts of supernovae. In particular, neutrinos will provide us with physical quantities, such as the radius and mass of protoneutron stars (PNS), which are the central engine of supernovae. This requires a theoretical model that connects observables such as neutrino luminosity and average energy with physical quantities. Here we show analytic solutions for the neutrino-light curve derived from the neutrino radiation transport equation by employing the diffusion approximation and the analytic density solution of the hydrostatic equation for a PNS. The neutrino luminosity and the average energy as functions of time are explicitly presented, with dependence on PNS mass, radius, the total energy of neutrinos, surface density, and opacity. The analytic solutions provide good representations of the numerical models from a few seconds after the explosion and allow a rough estimate of these physical quantities to be made from observational data.
\end{abstract}

\subjectindex{E11, E26, E32, E45}

\maketitle

\section{Introduction} \label{sec:intro}

Neutrinos are a guaranteed signal from a core-collapse supernova (SN) occurring in the nearby universe. We will observe many neutrinos when an SN appears inside the Milky Way with currently and future facilities. This is a strong point of neutrino observations compared to other observable signals. For instance, even inside the Milky Way, the optical emission would be absorbed by the dense dust if an SN happens near the Milky Way center, and the detectability of gravitational waves, whose amplitudes are highly uncertain and model dependent, is difficult to predict currently. For neutrinos, multiple current detectors, such as Super-Kamiokande \cite{abe16}, IceCube \cite{abba11}, and NO$\nu$A \cite{acer20}, and also future detectors expected within the next decade like Hyper-Kamiokande \cite{abe18}, JUNO \cite{an16}, and DUNE \cite{acci15} will be able to detect many neutrinos from the next nearby SNe. Preparations for observation are well underway.

On the other hand, not much progress has been made on the theory side. So far, only a few numerical studies for long-term protoneutron star (PNS) cooling associated with supernova explosions have been carried out using advanced technologies \cite{fisc10, hued10, robe12,naka13, suwa14,suwa19b,li20} (see Refs. \cite{sato87, burr88} for SN1987A and Ref. \cite{muel19} and references therein for short-term simulations, as well as observational predictions \cite{tam14}). These studies have just picked up a few progenitor models and presented a limited number of long-term evolution. To analyze the real data when an SN takes place, we need a systematic neutrino event rate evolution, i.e. neutrino-light curves. For this purpose, the detailed numerical simulations are not good enough for a parameter study, so we need simplified analytic formulae for neutrino-light curves. The same analogy applies in modeling the optical light curves of SNe, where the Arnett model \cite{arne82}, in which bolometric light curves are calculated by simply assuming photon diffusion in the expanding media, is widely used to extract physical parameters from observational data.
This model allows one to extract the mass of radioactive $^{56}$Ni, the mass of the ejecta, and the velocity of the ejecta with the bolometric light curves.
In this paper we give such analytic solutions for neutrino-light curves, which give simple and useful rules.

Although neutrino transport is a complex phenomenon dependent on the detailed structure of the ambient matter and the neutrino spectrum, we derive simple analytic formulae. We approximate the PNS's density structure by the Lane--Emden solution (Sect. \ref{sec:density}) and also approximate the neutrino transfer with the diffusion process (Sect. \ref{sec:transfer}). Combining two components (early-time and late-time solutions) gives good agreement with detailed numerical solutions. Based on the analytic formulae, we give a simple expression of the neutrino detection rate and positron energies, which are useful for the data analysis (Sect. \ref{sec:observables}). Some caveats are discussed in Sect. \ref{sec:disscussion} before the summary in Sect. \ref{sec:summary}.

\section{Density and temperature structure}
\label{sec:density}

We start from an analytic solution of the Lane--Emden equation for $n=1$ \cite{shap83},
\begin{equation}
\rho(\xi)=\rho_{\rm c}\frac{\sin\xi}{\xi},
\end{equation}
where $\rho_{\rm c}$ is the central density and $\xi$ the dimensionless radius. With the PNS mass $M_{\rm {PNS}}$ and radius $R_{\rm PNS}$, we can write $r=\alpha\xi$ with $\alpha=R_{\rm PNS}/\pi$ and $\rho_{\rm c}=M_{\rm PNS}/4\pi^2\alpha^3$. The surface of the PNS is at $\xi=\pi$. 
Note that $n=1$ corresponds to $\Gamma=2$, which gives a reasonable representation of nuclear-force-dominant regime, where $\Gamma$ is the adiabatic index (see, e.g. Ref. \cite{shap83}).

Assuming constant entropy over the whole PNS gives the temperature structure. The entropy per nucleon is \citep{beth90}
\begin{align}
s/k_{\rm B}=\frac{\pi^2}{2}\frac{k_{\rm B}T}{n_{\rm N}}\sum_{i=p,n}\frac{n_i}{\mu_i},
\end{align}
where $k_{\rm B}$ is the Boltzmann's constant, $T$ is the temperature, $n_{\rm N}$ is nucleon number density, $n_i$ is the number density of protons ($i=p$) and neutrons ($i=n$), and $\mu_i=(\hbar^2/2m)(3\pi^2 n_i)^{2/3}$ is the chemical potential of the nucleons, with $m$ being the mass of the nucleons. Here, we omit finite-temperature correction, since we are now interested in relatively cold PNSs. Since $n_{\rm N}=\rho/m$, we get
\begin{align}
k_{\rm B}T(\xi)
=&30\,{\rm MeV}
\left(\frac{M_{\rm PNS}}{1.4\,M_\odot}\right)^{2/3}
\left(\frac{R_{\rm PNS}}{10 \,{\rm km}}\right)^{-2}
\left(\frac{f(Y_e)}{0.630}\right)
\left(\frac{s}{1 \,k_{\rm B}\,{\rm baryon}^{-1}}\right)\left(\frac{\sin\xi}{\xi}\right)^{2/3},
\end{align}
where $f(Y_e)=[Y_e^{1/3}+(1-Y_e)^{1/3}]^{-1}$, with $Y_e$ being the electron fraction, i.e. $n_p=n_{\rm N}Y_e$ and $n_n=n_{\rm N}(1-Y_e)$. With $Y_e=0.1$; with $f(0.1)=0.699$ and $Y_e=0.5$, $f(0.5)=0.630$. 
Hereafter, we fix 
$f(Y_e)=0.630$
for simplicity because $f(Y_e)$ changes only slightly for the typical value of $Y_e$. With thermal energy per nucleon \citep{beth90}
\begin{align}
e_{\rm th}=\frac{\pi^2}{4}\frac{(k_{\rm B}T)^2}{n_{\rm N}}\sum_{i=p,n}\frac{n_i}{\mu_i},
\end{align}
the total thermal energy is given by
\begin{align}
E_{\rm th}
&=\int_0^{R_{\rm PNS}} 4\pi r^2 e_{\rm th}n_{\rm N} dr\nonumber\\
&=2.5\times10^{52}\,{\rm erg}
\left(\frac{s}{1k_{\rm B}\,{\rm baryon}^{-1}}\right)^2\left(\frac{M_{\rm PNS}}{1.4\,M_\odot}\right)^{5/3}
\left(\frac{R_{\rm PNS}}{10\,{\rm km}}\right)^{-2}.
\label{eq:Eth}
\end{align}
Note that the entropy outside the neutrinosphere might be different, but the position of the neutrinosphere is almost at the surface of the PNS for the timescale we are interested in; see Eq. \eqref{eq:pi-xi}. Thus, we assume that the entropy is constant over the whole PNS for simplicity.
For comparison, the total gravitational binding energy of this density structure is given as
\begin{align}
|W|
&=\int_0^{R_{\rm PNS}}4\pi GM_r\rho rdr\\
&=\frac{3}{4}\frac{GM_{\rm PNS}^2}{R_{\rm PNS}}\\
&=3.9\times 10^{53}\,{\rm erg}
\left(\frac{M_{\rm PNS}}{1.4\,M_\odot}\right)^2
\left(\frac{R_{\rm PNS}}{10\,{\rm km}}\right)^{-1},
\label{eq:Egrav}
\end{align}
which is larger than the case of constant density that gives $3.1\times 10^{53}$ erg.

\section{Neutrino transfer equation and solutions}
\label{sec:transfer}

The Boltzmann equation in spherical symmetry within $\mathcal{O}(v/c)$ is given by \cite{lind66}
\begin{align}
\frac{df}{cdt}
&+\mu\frac{\partial f}{\partial r}
+\left[\mu\left(\frac{d\ln\rho}{cdt}+\frac{3v}{cr}\right)+\frac{1}{r}\right](1-\mu^2)\frac{\partial f}{\partial \mu}
+\left[\mu^2\left(\frac{d\ln\rho}{cdt}+\frac{3v}{cr}\right)-\frac{v}{cr}\right]E\frac{\partial f}{\partial E}\nonumber\\
&=j(1-f)-\chi f+\frac{E^2}{c(hc)^3}
\left[(1-f)\int Rf'd\mu'-f\int R(1-f')d\mu'\right],
\label{eq:Boltzmann}
\end{align}
where $f=f(t,r,\mu,E)$ is the distribution function of neutrinos, $\mu$ is the cosine of the angle between the radial direction and the neutrino propagation, $E$ is the neutrino energy, $v$ is the fluid velocity with respect to the laboratory frame, $c$ is the speed of light, $j$ is the emissivity, $\chi$ is the absorptivity, and $R$ is the isoenergetic scattering kernel. We denote the Lagrangian time derivative in the comoving frame by $d/dt$. 
$f'$ refers to $f(t,r,\mu',E)$, where $\mu'$ is the angle cosine over which the integration is conducted.
Hereafter, we assume $v=0$ because we treat a static PNS as the background matter. 

Deep inside the PNS, the distribution function can be approximated by $f(t,r,\mu,E)\approx f^{(0)}(t,r,E)+\mu f^{(1)}(t,r,E)$.
Using this, the specific energy density and the specific flux are given by
\begin{align}
\varepsilon_E(t,r)&=\int_{4\pi}\frac{E^3}{(hc)^3}f d\Omega=\frac{2\pi E^3}{(hc)^3}\int fd\mu=\frac{4\pi E^3}{(hc)^3}f^{(0)},\\
F_E(t,r)&=\int_{4\pi}\frac{cE^3}{(hc)^3}f \mu d\Omega=\frac{2\pi c E^3}{(hc)^3}\int f\mu d\mu=\frac{4\pi}{3}\frac{cE^3}{(hc)^3}f^{(1)},
\end{align}
where $\Omega$ is a solid angle and $h$ is Planck's constant.
By introducing these into Eq. \eqref{eq:Boltzmann} and taking the zeroth and first angular moments, we get the following two equations:
 \begin{align}
\frac{\partial \varepsilon_E}{\partial t}+\frac{1}{r^2}\frac{\partial}{\partial r}\left(r^2F_E\right)&=
\frac{4\pi cE^3}{(hc)^3}j-(j+\chi)c\varepsilon_E,
\label{eq:0th-mon}\\
\frac{\partial F_E}{c\partial t}+\frac{c}{3}\frac{\partial \varepsilon_E}{\partial r}&=-(j+\chi+\phi)F_E,
\label{eq:1st-mon}
\end{align}
where $\phi:=\displaystyle\frac{E^2}{c(hc)^3}\int Rd\mu'$. 
Here, we also expand $f'(t,r,\mu',E)\approx f'^{(0)}(t,r,E)+\mu' f'^{(1)}(t,r,E)$ and assume $f^{(0)}=f'^{(0)}$ because we employ only elastic scattering for $R$.
By omitting the term $\partial F_E/\partial t$\footnote{In the diffusion limit, due to the short mean free path it becomes negligible compared to the source terms (see Sect. 80 in Ref. \cite{miha84}).} and introducing Eq. \eqref{eq:1st-mon} into Eq. \eqref{eq:0th-mon}, we get 
\begin{align}
\frac{\partial \varepsilon_E}{\partial t}-\frac{c}{3r^2}\frac{\partial}{\partial r}\left(\frac{r^2}{j+\chi+\phi}\frac{\partial \varepsilon_E}{\partial r}\right)&=\frac{4\pi cE^3}{(hc)^3}j-(j+\chi)c\varepsilon_E.
\label{eq:diffusioneq}
\end{align}
Since, for the local equilibrium state, the right-hand side of Eq. \eqref{eq:diffusioneq} vanishes, the equilibrium distribution function is given as $\varepsilon_E^{\mathrm{eq}}=\displaystyle\frac{4\pi E^3}{(hc)^3}\frac{j}{j+\chi}$. We rewrite Eq. \eqref{eq:diffusioneq} as
\begin{align}
\frac{\partial \varepsilon_E}{\partial t}-\frac{c}{3r^2}\frac{\partial}{\partial r}\left(\frac{r^2}{\kappa_t}\frac{\partial \varepsilon_E}{\partial r}\right)&=c\kappa_a(\varepsilon_E^{\rm eq}-\varepsilon_E).
\label{eq:diffusioneq2}
\end{align}
where we also rewrite the opacities with the variables $\kappa_t:=j+\chi+\phi$ and $\kappa_a:=j+\chi$. This gives the time evolution of the neutrino specific energy density.

Integrating over the energy gives 
\begin{align}
\frac{\partial \varepsilon}{\partial t}-\frac{c}{3r^2}\frac{\partial}{\partial r}\left(\frac{r^2}{\braket{\kappa_t}}\frac{\partial \varepsilon}{\partial r}\right)&=c(\braket{\kappa_a}^{\rm eq}\varepsilon^{\rm eq}-\braket{\kappa_a}\varepsilon),
\label{eq:diffusioneq3}
\end{align}
where 
\begin{align}
\varepsilon&:=\int_0^\infty \varepsilon_EdE\\
\varepsilon^{\rm eq}&:=\int_0^\infty \varepsilon_E^{\rm eq}dE=\frac{7}{16}aT^4,
\end{align}
with $a=\dfrac{8\pi^5 k_{\rm B}^4}{15 (hc)^3}=7.56\times10^{-15}$ erg cm$^{-3}$ K$^{-4}$ being the radiation constant. Here, we employ the Fermi--Dirac function without the chemical potential for the equilibrium spectrum of neutrinos as $\varepsilon_E^{\mathrm{eq}}=\displaystyle\frac{4\pi E^3}{(hc)^3}\frac{1}{1+e^{E/k_{\rm B}T}}$.
The opacities are expressed by the following mean:
\begin{align}
\frac{1}{\braket{\kappa_t}}&:=\frac{\displaystyle\int_0^\infty\frac{1}{\kappa_t}\frac{\partial \varepsilon_E}{\partial r}dE}{\displaystyle\int_0^\infty \frac{\partial \varepsilon_E}{\partial r}dE}=\frac{\displaystyle\int_0^\infty\frac{1}{\kappa_t}\frac{\partial \varepsilon_E}{\partial r}dE}{\displaystyle\frac{\partial \varepsilon}{\partial r}},\\
\braket{\kappa_a}&:=\frac{\displaystyle\int_0^\infty \kappa_a\varepsilon_E dE}{\varepsilon},\\
\braket{\kappa_a}^\mathrm{eq}&:=\frac{\displaystyle\int_0^\infty \kappa_a\varepsilon_E^{\mathrm{eq}}dE}{\varepsilon^{\mathrm{eq}}}.
\end{align}
The integrated flux is given by 
\begin{align}
F:=\int F_EdE=-\frac{c}{3}\int \frac{1}{\kappa_t}\frac{\partial \varepsilon_E}{\partial r}dE=-\frac{c}{3}\frac{1}{\braket{\kappa_t}}\frac{\partial \varepsilon}{\partial r}.
\end{align}

In the following, we assume that $1/\braket{\kappa_t}=1/\braket{\kappa_t}^{\mathrm{eq}}$ and $\braket{\kappa_a}=\braket{\kappa_a}^{\mathrm{eq}}$, since inside the PNS the spectrum is largely determined by the equilibrium state so that $\varepsilon_E\approx \varepsilon_E^{\mathrm{eq}}$. 
Then, taking $\kappa_t=\tilde \kappa_t(E/m_ec^2)^2$, $\kappa_a=\tilde \kappa_a(E/m_ec^2)^2$, with $m_e$ being the electron mass, we get
\begin{align}
\frac{1}{\braket{\kappa_t}^{\mathrm{eq}}}&=\frac{1}{2\tilde\kappa_t}\left(\frac{k_{\rm B}T}{m_ec^2}\right)^{-2}\frac{F_1}{F_3},
\label{eq:kappa_t}\\
\braket{\kappa_a}^{\mathrm{eq}}&=\tilde\kappa_a\left(\frac{k_{\rm B}T}{m_ec^2}\right)^{2}\frac{F_5}{F_3},
\label{eq:kappa_a}
\end{align}
where $F_n=\int_0^\infty x^n(1+e^x)^{-1}dx$ is a kind of complete Fermi--Dirac integral. $F_1=\pi^2/12\approx 0.822$, $F_2=3\zeta(3)/2\approx 1.80$, $F_3=7\pi^4/120\approx 5.68$, $F_4=45\zeta(5)/2\approx 23.3$, and $F_5=31\pi^6/252\approx 118$, where $\zeta(s)$ is the Riemann zeta function.

The opacities are given by Eqs. (21) and (22) of Ref. \cite{suwa19c} as
\begin{align}
    \tilde\kappa_a
    &=7.5\times 10^{-7}\,{\rm cm}^{-1}
    \left(\frac{\rho}{10^{14}\,{\rm g\,cm^{-3}}}\right)
    \label{eq:kappa_a2}
    ,\\
    \tilde\kappa_t
    &=4.0\beta\times 10^{-7}\,{\rm cm}^{-1}
    \left(\frac{\rho}{10^{14}\,{\rm g\,cm^{-3}}}\right).
\end{align}
Here, we fix $Y=0.5$ of Eq. \eqref{eq:kappa_a2} for simplicity (see Ref. \cite{suwa19c} for details).
$\beta$ is a boosting factor of the scattering due to the existence of heavy nuclei, which significantly amplify the scattering cross section by the coherent scattering \cite{brue85} in the PNS crust. $\beta\approx 3$ for free nucleons, in which the absorption is also taken into account.

With the thermal neutrino spectrum, we rewrite Eq. \eqref{eq:diffusioneq3} as
\begin{align}
\frac{\partial \varepsilon}{\partial t}-\frac{c}{3r^2}\frac{\partial}{\partial r}\left(\frac{r^2}{\braket{\kappa_t}^{\mathrm{eq}}}\frac{\partial \varepsilon}{\partial r}\right)&=c\braket{\kappa_a}^{\mathrm{eq}}(\varepsilon^{\mathrm{eq}}-\varepsilon).
\label{eq:diffeq}
\end{align}
From Eq. \eqref{eq:diffeq}, we can estimate the absorption and emission timescale, $\tau_{\rm a}$, and the diffusion timescale, $\tau_{\rm dif}$:
\begin{align}
\tau_{\rm a}&=\frac{1}{c\braket{\kappa_a}^{\mathrm{eq}}}
\approx 2.1\times 10^{-6}\,{\rm s}
\left(\frac{k_{\rm B}T}{m_ec^2}\right)^{-2}
\left(\frac{\rho}{10^{14}\,{\rm g\,cm^{-3}}}\right)^{-1},
\label{eq:tau_a}\\
\tau_{\rm dif}&\sim \frac{3R^2\braket{\kappa_t}^{\mathrm{eq}}}{c}
\approx
1.7\times 10^{-3}\,{\rm s}
\left(\frac{\beta}{3}\right)^{}
\left(\frac{R}{10\,{\rm km}}\right)^{2}
\left(\frac{k_{\rm B}T}{m_ec^2}\right)^{2}
\left(\frac{\rho}{10^{14}\,{\rm g\,cm^{-3}}}\right).
\label{eq:tau_d}
\end{align}
From them, if $(k_{\rm B}T/3\,{\rm MeV})^4\gtrsim(\rho/10^{11}\,{\rm g\,cm^{-3}})^{-2}(R/10\,{\rm km})^{-2}(\beta/3)^{-1}$, $\tau_{\rm dif}\gtrsim\tau_{\rm a}$, then $\varepsilon\approx \varepsilon^{\rm eq}$ is a good approximation.

By applying the approximation $\varepsilon=\varepsilon^{\rm eq}$ we have the following two equations:
\begin{align}
&\frac{\partial \varepsilon^{\rm eq}}{\partial t}+\frac{1}{r^2}\frac{\partial}{\partial r}\left(r^2 F^{\rm eq}\right)=0,\\
&F^{\rm eq}=-\frac{c}{3}\frac{1}{\braket{\kappa_t}^{\rm eq}}\frac{\partial \varepsilon^{\rm eq}}{\partial r}.\label{eq:Feq}
\end{align}
The boundary condition is given by $F^{\rm eq}={\rm const.}$ near the surface of a PNS, where the plane-parallel approximation can be applied. On replacing $r$ by the mean optical depth $\tau$, where
\begin{equation}
d\tau=-\braket{\kappa_t}^{\rm eq}dr,
\end{equation}
Eq. \eqref{eq:Feq} becomes
\begin{equation}
F^{\rm eq}=\frac{c}{3}\frac{d\varepsilon^{\rm eq}}{d\tau}.
\label{eq:F}
\end{equation}
Integrating Eq. \eqref{eq:F} with respect to $\tau$ gives 
\begin{align}
\varepsilon^{\rm eq}(\tau)=\varepsilon^{\rm eq}(\tau=0)\left(1+\frac{3}{2}\tau\right).
\end{align}
Here, we impose $F^{\rm eq}(\tau=0)=\dfrac{c}{2}\varepsilon^{\rm eq}(\tau=0)$ by assuming that neutrinos at the surface are isotropic along the outward directions and vanishing in the inward directions.
As $\varepsilon^{\rm eq}=\displaystyle\frac{7}{16}aT^4$, we have
\begin{align}
T(\tau)=T_0\left(1+\frac{3}{2}\tau\right)^{1/4},
\label{eq:T}
\end{align}
where $T_0$ is the surface temperature, i.e. $T_0:=T(\tau=0)$. It is related to the brightness temperature, which is given by $F=\dfrac{7}{16}\sigma T_{\rm br}^4$ with $\sigma=ac/4$ being the Stefan--Boltzmann constant, by
\begin{equation}
T_{\rm br}=2^{1/4}T_0.
\label{eq:Tbr}
\end{equation}
According to Eqs. \eqref{eq:T} and \eqref{eq:Tbr}, the brightness temperature becomes the same as the matter temperature at $\tau=2/3$, which is the definition of the {\it neutrinosphere} in the following. 

The radius of the neutrinosphere, $R_\nu$ is given by
\begin{align}
\frac{2}{3}&=\int_{R_\nu}^{R_{\rm PNS}}\braket{\kappa_t}^{\rm eq}dr\\
&=2\bar\kappa_t\frac{F_3}{F_1}\alpha g\rho_{\rm c}\left(\frac{k_{\rm B}T_c}{m_ec^2}\right)^2\int_{\xi_\nu}^\pi\left(\frac{\sin\xi}{\xi}\right)^{7/3}d\xi,
\label{eq:optical-depth}
\end{align}
where $\xi_\nu:=R_\nu/\alpha$, $\bar\kappa_t=\tilde\kappa_t/\rho$ and $T_{\rm c}:=T(\xi=0)$. We introduce a new parameter $g$ which accounts for the different structure of the PNS surface from the Lane--Emden solution with $n=1$.\footnote{The $g$ factor is typically smaller than unity since the polytropic index near the PNS surface is $n\approx 3$ (relativistic electrons are dominant), which leads to a steeper decline of the density than $n=1$. 
This is shown in Table \ref{tab:parameters}, in which $g=0.04$ --- 0.1 is used to fit the numerical solutions.}
The integral can be approximated as
\begin{equation}
\int_{\xi_\nu}^\pi\left(\frac{\sin\xi}{\xi}\right)^{7/3}d\xi
\approx
\frac{3}{10\pi^{7/3}}(\pi-\xi_\nu)^{10/3}.
\end{equation}
The error of this expression compared to the numerical integration is at most  17.4\% at $\xi_\nu\approx 1.2$.
Combining them gives
\begin{align}
\pi-\xi_\nu\approx 0.059
\left(\frac{M_{\rm PNS}}{1.4\,M_\odot}\right)^{-7/10}
\left(\frac{R_{\rm PNS}}{10\,{\rm km}}\right)^{9/5}
\left(\frac{g\beta}{3}\right)^{-3/10}
\left(\frac{s}{1\,k_{\rm B}\,{\rm baryon}^{-1}}\right)^{-3/5}.
\label{eq:pi-xi}
\end{align}
This tells us that the neutrinosphere is located near the PNS surface.

The neutrino luminosity is eventually given as
\begin{align}
L&=4\pi R_\nu^2F\\
&=4\pi (\alpha\xi_\nu)^2 \frac{7}{16}\sigma T(\xi_\nu)^4\\
&\approx \frac{7}{4}\pi^{1/3}\alpha^2 \sigma T_c^4(\pi-\xi_\nu)^{8/3}\\
&=1.2\times 10^{50}\,{\rm erg\,s^{-1}}
\left(\frac{M_{\rm PNS}}{1.4\,M_\odot}\right)^{4/5}
\left(\frac{R_{\rm PNS}}{10\,{\rm km}}\right)^{-6/5}
\left(\frac{g\beta}{3}\right)^{-4/5}
\left(\frac{s}{1\,k_{\rm B}\,{\rm baryon}^{-1}}\right)^{12/5},
\label{eq:luminosity}
\end{align}
where $\sin^{8/3}\xi_\nu/\xi_\nu^{2/3}\approx (\pi-\xi_\nu)^{8/3}/\pi^{2/3}$ for $\xi_\nu\approx \pi$, whose error is $\lesssim$3\% for $\pi-\xi_\nu\lesssim 0.6$, is used.
An error with a more precise estimate (9.6 $\times 10^{50}$ erg s$^{-1}$), which is given by a value of $\xi_\nu$ from Eq. \eqref{eq:optical-depth} solved numerically, is only $\sim 4\%$. For $s=0.1\, k_{\rm B}$ baryon$^{-1}$, the error becomes $\approx 20\%$ (approximate estimate gives 4.0$\times 10^{48}$ erg s$^{-1}$, while a numerical estimate gives 3.3$\times 10^{48}$ erg s$^{-1}$) because of a larger value of $\pi-\xi_\nu$ for a colder PNS.
Note that Eq. \eqref{eq:luminosity} gives the contribution for one flavor of neutrinos: that is, the total luminosity of all three flavors with neutrinos and antineutrinos is six times larger than Eq. \eqref{eq:luminosity}.

Next, we derive the time evolution of the neutrino luminosity. Since the PNS thermal energy is decreased by neutrino emission, 
\begin{align}
\frac{dE_{\rm th}}{dt}=-6L,
\end{align}
where we take into account all six types of neutrinos. By combining Eqs. \eqref{eq:Eth} and \eqref{eq:luminosity} we get 
\begin{align}
\frac{s}{1k_{\rm B}\,{\rm baryon}^{-1}}&=4.0\,
\left(\frac{M_{\rm PNS}}{1.4\,M_\odot}\right)^{13/6}
\left(\frac{R_{\rm PNS}}{10\,{\rm km}}\right)^{-2}
\left(\frac{g\beta}{3}\right)^{2}
\left(\frac{t+t_0}{100\,{\rm s}}\right)^{-5/2},
\end{align}
where $t$ is time and $t_0$ is the time origin, which gives the initial condition of the entropy. The corresponding neutrino luminosity is given as
\begin{align}
L=&3.3\times10^{51}\,{\rm erg\, s^{-1}}
\left(\frac{M_{\rm PNS}}{1.4\,M_\odot}\right)^{6}
\left(\frac{R_{\rm PNS}}{10\,{\rm km}}\right)^{-6}
\left(\frac{g\beta}{3}\right)^{4}
\left(\frac{t+t_0}{100\,{\rm s}}\right)^{-6}.
\label{eq:L_of_t}
\end{align}
Integrating $L$ over time and giving the total energy emitted by neutrinos $E_{\rm tot}$, $t_0$ is given by 
\begin{align}
t_0&=210\,{\rm s}
\left(\frac{M_{\rm PNS}}{1.4\,M_\odot}\right)^{6/5}
\left(\frac{R_{\rm PNS}}{10\,{\rm km}}\right)^{-6/5}
\left(\frac{g\beta}{3}\right)^{4/5}
\left(\frac{E_{\rm tot}}{10^{52}\,{\rm erg}}\right)^{-1/5}.
\label{eq:t0}
\end{align}
The average energy of neutrinos is given by
\begin{align}
\braket{E_\nu}:=
\dfrac{\varepsilon^{\rm eq}}{\displaystyle\int_0^\infty  (\varepsilon^{\rm eq}_E/E) dE}
=\frac{F_3}{F_2}k_{\rm B}T(\xi_\nu)
=16\, {\rm MeV}
\left(\frac{M_{\rm PNS}}{1.4M_\odot}\right)^{3/2}
\left(\frac{R_{\rm PNS}}{10\,{\rm km}}\right)^{-2}
\left(\frac{g\beta}{3}\right)
\left(\frac{t+t_0}{100\,{\rm s}}\right)^{-3/2}.
\label{eq:E_of_t}
\end{align}
Note that $\varepsilon_E^{\rm eq}/E$ is the specific number density.
For the early time (i.e. $t\ll t_0$), the luminosity and the average energy are 
\begin{align}
L(t=0)=&4.0\times10^{49}\,{\rm erg\, s^{-1}}
\left(\frac{M_{\rm PNS}}{1.4M_\odot}\right)^{-6/5}
\left(\frac{R_{\rm PNS}}{10\,{\rm km}}\right)^{6/5}
\left(\frac{g\beta}{3}\right)^{-4/5}
\left(\frac{E_{\rm tot}}{10^{52}\,{\rm erg}}\right)^{6/5},\\
\braket{E_\nu}(t=0)&=5.1{\rm MeV}
\left(\frac{M_{\rm PNS}}{1.4M_\odot}\right)^{-3/10}
\left(\frac{R_{\rm PNS}}{10\,{\rm km}}\right)^{-1/5}
\left(\frac{g\beta}{3}\right)^{-1/5}
\left(\frac{E_{\rm tot}}{10^{52}\,{\rm erg}}\right)^{3/10}.
\end{align}

In the model, there are five parameters: the PNS mass $M_{\rm PNS}$, the PNS radius $R_{\rm PNS}$, the density correction factor $g$, the opacity boosting factor by coherent scattering $\beta$, and the total energy emitted by all flavors of neutrinos $E_{\rm tot}$. Note that the boosting factor $\beta$ is time dependent because the heavy nuclei in the crust are absent for the early phase and appear later once the temperature decreases below the Coulomb energy of the lattice structure \cite{suwa14}. Therefore, we propose a two-component model to reproduce numerical models of neutrino-light curves. The first component represents the early time without coherent scattering ($\beta=3$), and the second component represents the late time with the opacity boost by the coherent scattering ($\beta\gg 1$). The neutrino luminosity is given by the total luminosity of the two components, $L_1+L_2$, and the average energy is estimated by the harmonic mean, $\dfrac{L_1+L_2}{L_1/\braket{E_1}+L_2/\braket{E_2}}$, where $L_i$ and $\braket{E_i}$ give the luminosity and average energy of the $i$-th component.

\begin{figure}[htbp]
\centering
\includegraphics[width=0.7\textwidth]{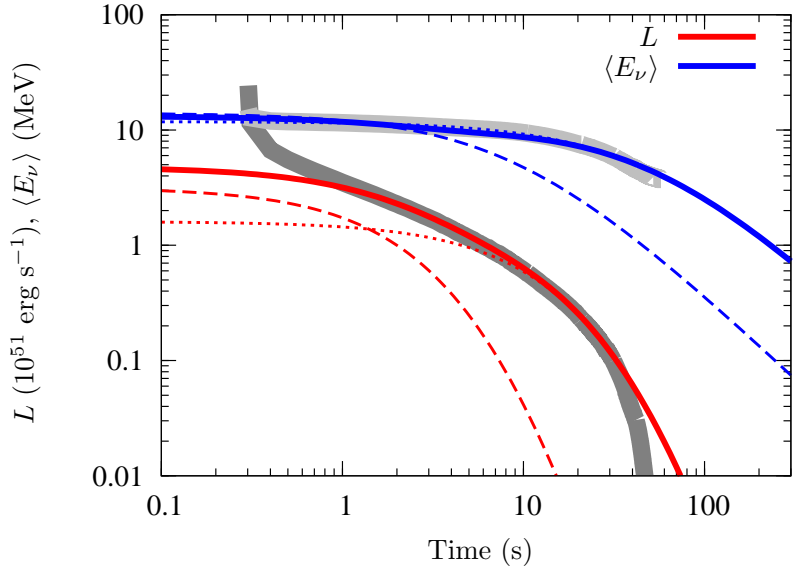}
\caption{Luminosity (red) and average energy (blue) evolution for one neutrino flavor. The first component is a model with $\beta=3$ and $E_{\rm tot}=4\times 10^{52}$ erg: the second component is a model with $\beta=40$ and $E_{\rm tot}=1\times 10^{53}$ erg. For both components, $M_{\rm PNS}=1.5\,M_\odot$, $R_{\rm PNS}=12$ km, and $g=0.04$. The gray lines are the luminosity and average energy of $\bar\nu_e$ in the model 147S of Ref. \cite{suwa19b}. }
\label{fig:analytic_model}
\end{figure}

Figure \ref{fig:analytic_model} shows a comparison of the analytic model given here (colored lines) and the numerical model 147S presented in Ref. \cite{suwa19b} (gray lines), which is a numerical solution of a PNS cooling calculation that solves the neutrino transfer equation with a nuclear-physics-based equation of state as well as the general relativistic hydrostatic equation. For the analytic model, we employ the early-time solution (dashed lines) and the late-time solution (dotted lines). The early-time solution indicates the cooling curve without the solid crust composed of heavy nuclei (i.e. low $\beta$), while the late-time solution includes it  (i.e. high $\beta$). The solid red line is the total luminosity of the early-time and late-time solutions, and the solid blue line is the harmonic mean of the two average energies. The general profiles of the detailed numerical solutions are reproduced well by the simple analytic solutions presented in this paper. In the very early phase ($t\lesssim 1$ s), the PNS contracts so that the gravitational energy converts to additional neutrino emission. In the very late phase ($\gtrsim 30$ s), the approximations (e.g. the thermal spectrum, a constant $g$ factor) may break down.
This shows that the analytic solutions are valid from $\sim 1$ to several tens of seconds.

\begin{table}[]
    \centering
    \caption{Model parameters reproducing numerical solutions presented in Ref. \cite{suwa19b}.}
    \begin{tabular}{cccccccc}
    Model & $M_{\rm PNS}$ ($M_\odot$) & $R_{\rm PNS}$ (km) & $g$ & $\beta_1$ & $E_{\rm tot,1}$ ($10^{52}$ erg) & $\beta_2$ & $E_{\rm tot,2}$ ($10^{52}$ erg)\\
    \hline
    147S & 1.5 & 12 & 0.04 & 3 & 4.0 & 40 & 10 \\
    M1L  & 1.3 & 11 & 0.04 & 3 & 2.5 & 25 & 5.0 \\
    M1H  & 1.3 & 11 & 0.04 & 3 & 2.5 & 30 & 9.0  \\
    M2L  & 2.3 & 13 & 0.1 & 3 & 8.0 & 30 & 22 \\
    M2H  & 2.3 & 13 & 0.1 & 3 & 11 & 40 & 35
    \end{tabular}
    \label{tab:parameters}
\end{table}

Model 147S is just an example in a series of models presented in Ref. \cite{suwa19b}; we also fit the other models and summarize the fitting parameters in Table \ref{tab:parameters}.
These parameter sets can be used for mock data production to study detector responses for the next nearby supernovae. They are also useful to complete the neutrino-light curve with the detailed (multidimensional) hydrodynamics simulations, which are typically calculated up to $\sim 1$ s, by connecting the numerical data with the analytic formula for the PNS cooling phase.
Here, we compare the PNS mass and radius obtained from the analytic solutions with those of the numerical models. The PNS mass is consistent with the numerical model (147S has 1.47 $M_\odot$, M1L and M1H have 1.29 $M_\odot$, and M2L and M2H have 2.35 $M_\odot$ for the baryonic mass) due to degeneracy with other parameters, but the PNS radius differs from the numerical model (147S has 14.3 km, M1L and M1H have 14.4 km, and M2L and M2H have 13.4 km). This may be because the analytic solutions apply the Lane--Emden solution to the entire region, and is based on Newtonian gravity. The resolution of these deviations is beyond the scope of this paper, but we will improve them in future work.

Although we demonstrated that fitting by two components works quite well, we limit our discussion to the single-component model fitting the experimental data to extract physical quantities in the next section. Fitting will be elaborated for two components in future study.

\section{Observables and parameter extraction}
\label{sec:observables}

With the simple analytic formula, we can estimate the event rate evolution with a water-Cherenkov detector like Super-Kamiokande. 
Here, we focus only on the second component of the previous section, which dominates the late-time properties and is useful for extracting physical parameters.
The event rate is approximately given by the total number of protons in the detector, the number of anti-electron-type neutrinos coming into the detector, and the cross section of inverse beta decay that is the main interaction capturing neutrinos. The event rate is given by 
\begin{align}
    \mathcal{R}&\approx
    \frac{2}{18}\frac{M_{\rm det}}{m}
    \frac{L}{4\pi D^2\braket{E_\nu}}
    \braket{\sigma}\\
    &\approx 210\,{\rm s^{-1}}
    \left(\frac{M_{\rm det}}{32.5\,{\rm kton}}\right)
    \left(\frac{D}{10\,{\rm kpc}}\right)^{-2}
    \left(\frac{L}{10^{51}\,{\rm erg\,s^{-1}}}\right)
    \left(\frac{\braket{E_\nu}}{15\,{\rm MeV}}\right),
\end{align}
where 
$M_{\rm det}$ is the detector mass (32.5 kton corresponds to the entire volume of the inner tank of Super-Kamiokande), $m$ is the nucleon mass, $D$ is the distance between an SN and the Earth, and $\sigma$ is the cross section of inverse beta decay ($p+\bar\nu_e\to n+e^{+}$). For the cross section, we use $\sigma(E)=\sigma_0(E/{\rm MeV})^2$ with $\sigma_0=9.4\times 10^{-44}\,{\rm cm}^2$ \cite{raff96} and, by assuming the thermal spectrum,
$\braket{\sigma}=\int \sigma(E)(\varepsilon_E^{\rm eq}/E)dE/\int (\varepsilon_E^{\rm eq}/E)dE=(F_4/F_2)\sigma_0(k_{\rm B}T/{\rm MeV})^2=(F_2F_4/F_3^2)\sigma_0(\braket{E_\nu}/{\rm MeV})^2$. 
Since the factor $2/18$ is for counting the hydrogen number in water molecules, 2$M_{\rm det}/18m$ is the total number of hydrogen in the detector. 
Introducing Eqs. \eqref{eq:L_of_t} and \eqref{eq:E_of_t}, we get
\begin{align}
\mathcal{R}&\approx
720\,{\rm s^{-1}}
\left(\frac{M_{\rm det}}{32.5\,{\rm kton}}\right)
\left(\frac{D}{10\,{\rm kpc}}\right)^{-2}
\left(\frac{M_{\rm PNS}}{1.4\,M_\odot}\right)^{15/2}
\left(\frac{R_{\rm PNS}}{10\,{\rm km}}\right)^{-8}
\left(\frac{g\beta}{3}\right)^{5}
\left(\frac{t+t_0}{100\,{\rm s}}\right)^{-15/2}.
\label{eq:detection_rate}
\end{align}
The average energy of positrons is given by
\begin{align}
E_{e^+}\approx\dfrac{\displaystyle\int_0^\infty \sigma(E) \varepsilon^{\rm eq}_E dE}{\displaystyle\int_0^\infty \sigma(E) (\varepsilon^{\rm eq}_E/E) dE}
&=\frac{F_5}{F_4}k_{\rm B}T(\xi_\nu)=\frac{F_5F_2}{F_4F_3}\braket{E_\nu}\nonumber\\
&=25\, {\rm MeV}
\left(\frac{M_{\rm PNS}}{1.4\,M_\odot}\right)^{3/2}
\left(\frac{R_{\rm PNS}}{10\,{\rm km}}\right)^{-2}
\left(\frac{g\beta}{3}\right)
\left(\frac{t+t_0}{100\,{\rm s}}\right)^{-3/2}.
\label{eq:detection_energy}
\end{align}
Note that the positron energy given above is valid only when the typical energy of positrons is sufficiently higher than the threshold energy of the data analysis, typically 5 MeV (see Sect. 4.2  of Ref. \cite{suwa19b}).

Once we detect neutrinos from the next nearby SN, these formulae can be applied to narrow down the parameter space, 
which would give a starting point for more detailed calculations. For instance, by dividing Eq. \eqref{eq:detection_rate} by the fifth power of Eq. \eqref{eq:detection_energy}, one finds
\begin{align}
R_{\rm PNS}
=10\,{\rm km}
\left(\frac{\mathcal{R}}{720\,{\rm s^{-1}}}\right)^{1/2}
\left(\frac{E_{e^+}}{25\, {\rm MeV}}\right)^{-5/2}
\left(\frac{M_{\rm det}}{32.5\,{\rm kton}}\right)^{-1/2}
\left(\frac{D}{10\,{\rm kpc}}\right),
\end{align}
which is time independent. Next, by taking maximum of $t\mathcal{R}$ one finds
\begin{align}
    \frac{d}{dt}(t\mathcal{R})=0 \rightarrow t=\frac{2}{13}t_0,
\end{align}
which gives $t_0$. Since $M_{\rm det}$ is given by experiment and $D$ would be measured by the optical or infrared observations, the unknowns are $M_{\rm PNS}$, $g$, and $\beta$. Unfortunately, at the current moment they are degenerate and the combination $M_{\rm PNS}(g\beta)^{2/3}$ is only measurable.
The degeneracy would be resolved by combining all the available data, for instance the total energy emitted by neutrinos gives information on $GM_{\rm PNS}^2/R_{\rm PNS}$ --- see Eq. \eqref{eq:Egrav}.

Also, a consistency relation for the analytic formulae is given by 
\begin{align}
\frac{\mathcal{R}\ddot{\mathcal{R}}}{\dot{\mathcal{R}}^2}=\frac{17}{15},
\end{align}
where a dot denotes the time derivative. By measuring it the model consistency can be tested.

\section{Discussion}
\label{sec:disscussion}

We discuss here caveats for our drastic assumptions to make the formulae simple. The assumptions of the analytic solutions are the following.
We employ the Lane--Emden solution with $n=1$ for the density and temperature profiles of neutron stars and assume constant entropy profiles. For the neutrino transfer equation, we employ thermal equilibrium (i.e., the Fermi--Dirac distribution). Also, spherical symmetry is applied.

These assumptions lead to caveats as follows:
First, the density profiles are dependent on the nuclear equation of state so that we need a more realistic equation of state for the detailed evolution, which makes the analytic treatment difficult. 
Second, the entropy profile is not constant just after the explosion, but in the late phase, the neutrino diffusion produces almost constant entropy profiles \cite{naka13,suwa14}. Therefore, at least for the late time, a constant entropy profile is not a wrong assumption.
Third, PNS convection may change the neutrino luminosity in the early time, but it becomes weak in the late time so that our analytic formula would not be affected. 
Fourth, the neutrino spectrum is not purely thermal when the temperature gets low, and the diffusion timescale becomes shorter than the emission/absorption timescales in the very late time. To give a more realistic luminosity and spectrum, we need to solve the multi-energy neutrino transfer equation, which is the next step. 
Lastly, neutrino oscillation is not included in this study. But, it is expected that, in the late time, all flavors of neutrinos have similar luminosity and spectrum so that neutrino oscillation does not change the analytic solutions \cite{suwa19b}.

\section{Summary}
\label{sec:summary}

We have derived analytic solutions of the neutrino radiation transfer equation in protoneutron star cooling after core-collapse supernovae. The luminosity evolution is given by Eq. \eqref{eq:L_of_t} and the average energy evolution by Eq.  \eqref{eq:E_of_t}. The evolution of the detection rate and the positron energies for water-Cherenkov detectors are also given by Eqs. \eqref{eq:detection_rate} and \eqref{eq:detection_energy}. The timescale in these equations is Eq.  \eqref{eq:t0}. With these equations, the neutrino emissions for the late time, in particular $\gtrsim 1$ s after the explosion, are given.

The analytic solutions presented in this paper are the very first step of analytic expression of detectable neutrino-light curves; the formulae will be updated to give more useful expressions for the next galactic supernova in forthcoming papers.

\section*{Acknowledgments}

This work was supported by Grants-in-Aid for Scientific Research (19K03837, 19K23435, 20H00174, 20H01904, 20H01905, 20K03973) and Grants-in-Aid for Scientific Research on Innovative Areas (17H06365, 18H04586, 18H05437, 19H05811, 20H04747) from the Ministry of Education, Culture, Sports, Science and Technology (MEXT), Japan.
This work was supported by 
MEXT as Program for Promoting Researches on the Supercomputer Fugaku,
``Toward a unified view of the universe: from large scale structures to planets.''
This work was partially carried out by the joint research program of the Institute for Cosmic Ray Research (ICRR), The University of Tokyo.


%

\end{document}